\documentclass{article}
\usepackage[utf8]{inputenc}
\usepackage{multirow}
\usepackage{url}
\usepackage{rotating}
\newcommand{\up}{\vspace*{-15pt}}
\usepackage[hide]{chato-notes} 

\title{Quantified Self Meets Social Media: Sharing of Weight Updates on Twitter}

\author{
Yafei Wang\\
      \texttt{The Pennsylvania State University}\\
      \texttt{University Park}\\
      \texttt{PA 16802, USA}\\
      \texttt{yxw184@ist.psu.edu}
      \and
Ingmar Weber \\ 
      \texttt{Qatar Computing Research Institute, HBKU}\\
      \texttt{Tornado Tower, P.O.\ 5825}
      \texttt{Doha, Qatar}\\
      \texttt{iweber@qf.org.qa}
      \and
Prasenjit Mitra\\
      \texttt{Qatar Computing Research Institute, HBKU}\\
      \texttt{Tornado Tower, P.O.\ 5825}
      \texttt{Doha, Qatar}\\
      \texttt{pmitra@qf.org.qa}
}

\begin{document}

\maketitle

\begin{abstract}

An increasing number of people use wearables and other smart devices to quantify various health conditions, ranging from sleep patterns, to body weight, to heart rates. Of these ``Quantified Selfs'' many choose to openly share their data via online social networks such as Twitter and Facebook. In this study, we use data for users who have chosen to connect their smart scales to Twitter, providing both a reliable time series of their body weight, as well as insights into their social surroundings and general online behavior. Concretely, we look at which social media features are predictive of physical status, such as body weight at the individual level, and activity patterns at the population level. We show that it is possible to predict an individual's weight using their online social behaviors, such as their self-description and tweets. Weekly and monthly patterns of quantified-self behaviors are also discovered. These findings could contribute to building models to monitor public health and to have more customized personal training interventions.

While there are many studies using either quantified self or social media data in isolation, this is one of the few that \emph{combines} the two data sources and, to the best of our knowledge, the only one that uses public data.

\end{abstract}

\todo[Ingmar]{@Yafei: If there is any space issue, then remove the concepts and the keywords from the first page, but keep them for the submission system.}

\section{Introduction}

During the last couple of years, the number of users who use ``Quantified Self'' (QS) health tracking devices has continuously increased \footnote{\url{http://nuviun.com/digital-health/quantified-self}}. A survey of 1,262 U.S.\ adult consumers conducted in December 2014 found that 31\%  use a QS tool to track their health and fitness.\footnote{\url{http://quantifiedself.com/docs/RocketFuel_Quantified_Self_Research.pdf}} 
In lockstep with this proliferation of QS tools, research output on all related aspects has also seen a dramatic increase. 
Google Scholar lists 74 publications with ``Quantified Self'' in the title for \emph{all} years up to and including 2013.\footnote{\url{https://scholar.google.com/scholar?q=intitle\%3A"quantified+self"&as_yhi=2013}. Last accessed on Jan 8, 2016.} 
However, since 2014 alone, already 123 publications matching this criteria have been indexed.\footnote{\url{https://scholar.google.com/scholar?q=intitle\%3A"quantified+self"&as_ylo=2014}. Last accessed on Jan 8, 2016.} 
Most of this research, however, looks at QS data in isolation, separately from other data one might obtain for a user.

In this paper we present results from an attempt to \emph{combine} QS data with social media data. This link between two data sources is made possible as more and more users publicly share the QS data they generate. For our study, we use data from users who have chosen to connect their smart scale to their public Twitter stream.

\begin{figure}
  \centering
  \includegraphics[width=1\columnwidth]{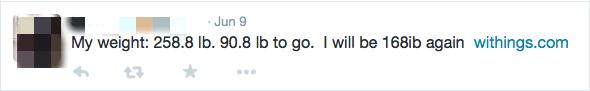}
  \caption{An anonymised example of a self-quantified English tweet generated by Withings' smart scale.}~\label{fig:engSample}
 \up
\end{figure}

Concretely, we analyze data for users who have opted in to connect their internet-enabled Withings Smart Body Analyzer\footnote{\url{http://www2.withings.com/us/en/products/smart-body-analyzer}} to their Twitter account. 
Figure~\ref{fig:engSample} shows an anonymised example tweet.
We analyze data for 897 Twitter users who (i) not only have auto-generated fitness tweets from Withings, and apps such as Fitbit, or Nike, but (ii) also have  ``normal'' tweets not generated by fitness apps. 

We are interested to see if there is a link between these two data sources and, say, if a user's weight can be inferred from their general social media behavior. Do users with a larger body weight somehow tweet differently? If a link can be found, then this opens up new opportunities as it hints at the ``clinical relevance'' of social media data. In particular, we envision that social media could be used as one building block, together with QS data and electronic health records, to devise more personalized, holistic interventions that take a user's life style into account \cite{haddadietal15ichi}. For example, a doctor could be provided with information about the personality of the individual from their tweets or the influences on the individual from their social circles that need to be taken into account (and, in some cases, overcome) while understanding and prescribing diet and exercise interventions.

Combining QS data and social media data also helps to overcome one frequent shortcoming of health-related studies: a lack of individual-level ground truth. While county-level health statistics, such as obesity rates, are readily available, it is much harder to obtain a set of Twitter users with a known weight, ideally also traceable over time. Despite the obvious limitations due to selection bias, users who link their QS data to their public social media account still provide a valuable data set.

At the same time, there are technical challenges that need to be addressed for a successful data fusion. Social media data is famously noisy due to internet lingo, spam and bots, and data incompleteness resulting from API limitations. The QS data also has its share of issues as users share their scale with friends\footnote{We even observed an instance where, apparently, the weight of a \emph{cat} was recorded.} or they might weigh themselves both before and after eating a large meal or going to the rest room. The combined data creates additional challenges due to its heterogenous nature: a textual stream (and more) from normal tweets, and a time series of weight data from QS tweets. We help address some of these issues by describing a method to remove implausible weigh-in data points.

Apart from technical challenges, the ethical challenges are at least as daunting. Though legally ``public'' data, tweets still often contain information that many would consider ``private''. This is arguably due to a misconception of the perceived audience (i.e., a user's Twitter followers) vs.\ the actual audience (i.e., data scientists around the world). Such concerns are amplified for the domain of medical data.

\todo[Ingmar]{@Yafei: make sure that the R\^2 result stated is the latest.}
\note[Yafei]{Changed it to the latest result.}

We believe that tackling these challenges is well worth the effort, especially as our initial results are promising. Concretely, we find that social media data can predict a user's average weight with an R-squared around $.25$. We also find that the QS data collected through the Twitter stream is valuable by itself for population-level health analysis. For example, it lets us paint a picture of weight transition across the year -- yes, it goes up over Christmas and New Year -- and of ``dieting morale'' across the week -- users weigh themselves most often on weekends which, ironically, is when they are least likely to do something about their weight.  
Given that we would not have been able to collect this data otherwise, this is a further advantage of looking at the intersection between QS and social media.

\section{Related Work}

To the best of our knowledge, there is very little prior work that \emph{combines} social media and QS data at the level of the individual. The vision described by Estrin~\cite{estrin14cacm} definitely includes a combination of data sources but no study on such data seems to have been performed to date. 
Vickey and Breslin \cite{vickeybreslin12aaaiss} report a system-level study of how fitness app data is shared on Twitter, but they do not include a user-level study that links a user's normal Twitter data with their fitness data. The StudentLife Project\footnote{\url{http://studentlife.cs.dartmouth.edu/}} \cite{wangetal14ubicomp} uses a mobile phone app to collect detailed activity data which is linked to academic performance. This data also includes Facebook profiles, though these are not part of the publicly shared data set. 

Concerning work more closely related to obesity and weight loss, studies using social media typically take a population-level, public health approach. Culotta~\cite{culotta14chi} used geo-tagged tweets and Abbar et al.\ \cite{abbaretal15chi} used food-related tweets to predict geographical differences in obesity and diabetes. Though including ``normal life'' in their analysis, they use county-level data as ``ground truth'' for obesity. By using quantified self data, we can obtain weight-related information at the individual level.

There is also a body of work that studies specialized social media, such as online weight loss forums \cite{lietal14group}. Particular attention has been given to predicting weight loss from interactions in the social network. 
Chomutare et al.~\cite{chomutareetal14cbms} showed that high levels of activity in online obesity communities and being connected to several disparate sub-communities were both predictive of weight loss.  They observed that the network structure properties were more useful in predicting weight loss than the biographical information associated with the users.
Li et al.~\cite{lietal15imds} take this a step further by studying the problem of recommending a ``good'' friend within the context of a weight loss app. 
Brindal et al.~\cite{brindal2012jmir} showed that the inclusion of a social networking platform did not have additive effects with respect to weight loss or retention.  However, these inclusions resulted in patients using their weight loss system for a longer duration.  However, in their experiments, greater use of a weight tracker tool resulted in greater loss. Though our work only looks at the combination of social media and QS for \emph{data collection}, their work provides evidence for benefits for \emph{health interventions} as well.

\section{Data Collection}\label{sec:Data:Description}
To collect our data, we use the Twitter Streaming API\footnote{\url{https://dev.twitter.com/streaming/overview}} for three weeks in Oct 2015 collecting tweets containing keywords, ``lb'' or ``kg''. Note that this broad pattern captures data both weight-related QS tweets and other tweets in several languages.  We also use the Topsy API\footnote{\url{https://otter.topsy.com/search.json?q=kg+lb}, no longer supported.} to gather all obtainable historical tweets containing these keywords.
These tweets were then post-filtered such that 
only tweets being generated by ``WiTwit'' were kept.\footnote{``WiTwit'' is the ``source'' field used in Tweets generated by WiThings' smart scale.} 
For each unique user who generated these tweets, we obtained (i) (up to) 3,200 of their most recent tweets, (ii) their self-generated profile known as ``bio'', (iii) the lists of their friends and followers, and (iv) the bios of their friends and followers. 

Inspecting the data, we observed that a large fraction of users \emph{only} tweeted their weight or other automatic fitness tweets. These specially created accounts, potentially as a sort of personal fitness log, were not of interest for us as, apart from a time series of weigh-ins, there was no other social media data to better understand the user. We therefore imposed an additional filter by requiring (i) each user to have at least 10 ``normal'' tweets \emph{not} automatically generated by one of WiTwit, or FitBit \footnote{The full list of apps considered for this is MyFitnessPal, Fitbit, Withings, Lose It! and Nike.}. 
and (ii) having at least ten weigh-in tweets automatically generated by WiTwit.

As we additionally wanted to make sure that users, at least potentially, have social interactions on Twitter, we further required all users to have at least 50 friends and followers, for individual analysis. The cutoff of 50 was chosen based on manual inspection. 
For example, a specific user with 60 friends and 41 followers, just below the cutoff, only published 269 tweets, including only 30 normal ones. Another user with 657 friends and 55 followers, just above the cutoff, published 837 tweets, including 746 normal ones. In total, we excluded 467 users, and with 430 users remaining after filtering for social interactions.

\subsection{Identification of Fitness Tweets Generated By Fitness Apps}
As mentioned earlier, a large fraction of users in our data set had additional automatically-generated fitness  tweets, apart from the ones from Withings. We collected these tweets separately as they hold additional, valuable QS information. In order to identify these  tweets, we check the source field of each individual tweet. The source field indicates the tool used to post the Tweet. For auto-generated tweets, the source field provides the name and the URL of the corresponding app, such as WiTwit, Runkeeper, or Fitbit. Table~\ref{FitnessPattern} shows the patterns of automatic fitness tweets we have used in this paper. 300 out of 897 users in our dataset have at least one automatic fitness tweet.

\todo[Ingmar]{@Yafei: above we mentioned that only 430 users remain. But here the statistics is with reference to 897}
\note[Yafei]{897 users are users without social interaction cleaning.}

\begin{table}[]
\centering
\begin{tabular}{|l|l|}
\hline
Type & Patterns \\ \hline
Original Weight Loss & WiTwit \\ \hline
Other Weight Loss & Lose It!, SimpleWeight\\ \hline
Fitness & \begin{tabular}{@{}l@{}}
RunKeeper, Fitbit, Nike \\ 
Runmeter, Runtastic, Nike+ GPS, \\  
iSmoothRun 
\end{tabular}
\\
\hline
\end{tabular}
\caption{List of patterns identifying automatic fitness tweets in the ``source'' field of a tweet's JSON file. 
}
\label{FitnessPattern}
\up
\end{table}

\section{Linking Social Media Behavior and Quantified-Self Data at the Individual Level}
In this section, we utilize users' online social activities to predict their body weight. Their  weight is measured by Withing scales and we assign each person the average of all of their recorded weigh-ins as their reference weight. Two types of Twitter data sources are utilized as features to predict this weight:  (i) their self-description (also known as bio) and (ii) their tweets. All non-English content was translated to English using Google's machine translation \footnote{\url{https://translate.google.com}}.

Upon inspection of the data,  we found that some users share their Withing scales with their family members. To detect and clean such weigh-in series generated by multiple people, we apply a formula for ``plausible weight transitions'': for a given user, a weight transition from weight w(i) to w(i+1) [in pounds] between days d(i) and d(i+1) is ``plausible'' if $|w(i) - w(i+1)|  \le  4 + | d(i) - d(i+1)|$. In words, we allow for up to 4lb of weight fluctuation within one day and 1lb for each day passed. Note that 1lb of body fat is roughly equivalent to 3,500kcal. Though larger fluctuations are possible, especially due to excessively storing or losing liquid, we decided to err on the side of caution, rather than including too much erroneous data. Users with more than three plausibility violations were excluded. 
In addition, we observed that some users reported suspiciously low or high some weights such as 12 lbs or 400 lbs. Therefore, users whose average weight is either smaller than 100 lbs or larger than 300 are treated as outliers and excluded from our analysis. 
As a point of reference, the average of the individual average weights was 178.4lb. Note that this is very close to the 2012 average weight of adults in North America of 177.9lb\footnote{\url{https://en.wikipedia.org/wiki/Human_body_weight#By_region}}, indicating that our data might not be as odd and biased as one might imagine.
After applying the data filtering explained above, we use the remaining 391 users to build a prediction model.

\begin{table}[]
\centering
\begin{tabular}{|l|l|l|l|}
\hline
Feature Name         & Coef. & Feature & Coef. \\ \hline
Bio\_LIWC\_ppron    & 91.44 & Tweet\_LIWC\_auxverb & -20.01 \\ \hline
Bio\_LIWC\_affect   & 90.97 & Tweet\_LIWC\_verb    & -19.55 \\ \hline
Bio\_LIWC\_pronoun  & 90.71 & Tweet\_LIWC\_social  & -18.10 \\ \hline
Tweet\_LIWC\_feel   & 90.05 & Tweet\_LIWC\_number  & -16.82 \\ \hline
Bio\_LIWC\_social   & 89.57 & Tweet\_LIWC\_present & -16.40 \\ \hline
Tweet\_LIWC\_ingest & 86.24 & Tweet\_LIWC\_past    & -15.98 \\ \hline
Bio\_LIWC\_present  & 79.47 & Tweet\_LIWC\_article & -15.56 \\ \hline
Bio\_LIWC\_auxverb  & 76.12 & Tweet\_LIWC\_conj    & -15.50 \\ \hline
Bio\_LIWC\_verb     & 76.09 & Tweet\_LIWC\_adverb  & -15.26 \\ \hline
Tweet\_LIWC\_shehe  & 75.62 & Tweet\_LIWC\_excl    & -15.00 \\ \hline
Bio\_LIWC\_incl     & 72.23 & Tweet\_LIWC\_funct   & -14.64 \\ \hline
Bio\_LIWC\_cogmech  & 71.20 & Tweet\_LIWC\_insight & -13.91 \\ \hline
Bio\_LIWC\_article  & 69.83 & Tweet\_LIWC\_tentat  & -13.48 \\ \hline
Bio\_LIWC\_conj     & 66.49 & Tweet\_LIWC\_you     & -11.00 \\ \hline
Tweet\_LIWC\_bio    & 65.97 & Tweet\_LIWC\_discrep & -10.40 \\ 
\hline
\end{tabular}
\caption{A Support Vector Machine model with linear kernel for predicting a person's average weight using their tweets and self-description. The top-15 features for each direction are shown here.}
\label{Model}
\up
\end{table}

In order to capture and summarize a user's social media content, we utilize two existing dictionaries that have undergone psychometric validation. The first is the Linguistic Inquiry and Word Count (LIWC) dictionary\footnote{We use the LIWC2007 dictionary in this paper.} \cite{pennebaker07liwc} with 64 categories, and the second is the PERMA\footnote{PERMA is a mnemonic for Positive emotion, Engagement, Relationships, Meaning, and Achievement --- 
the five elements of well-being.} dictionary \cite{seligman12perma} with ten categories. 
Both dictionaries map terms to a set of categories such as ``social'', ``health'' and ``body'' in the case of LIWC \cite{pennebaker07liwc}, or ``positive emotion'', ``engagement'' and ``meaning'' in the case of PERMA. For example, PERMA maps the term ``distract'' to ``negative emotion''; and LIWC maps the term ``brother'' to ``social''. We applied this mapping both to a user's normal tweets\footnote{Automatically generated fitness tweets and weigh-in tweets are excluded.} and their bio. For boosting the model performance, we also add Bag of Word features.

In order to quantify and interpret the effects of different indicators, we fit a support vector machine model with a linear kernel to predict their personal weight at the individual level. All the social activity features (except their actual weight) have been linear max-min scaled to [0,1]. The top 15 indicators of the support vector machine model with linear kernel for each direction are shown in Table~\ref{Model}.

Given the model in Table~\ref{Model}, it is worth looking at which Twitter features are most predictive of a person's weight.
People with higher actual weight mentioned more ingest words (\emph{Tweet\_LIWC \_ingest}), such as food, dish, and eat, in their self description. This might suggest that people who publicly express their love for food have a higher probability to be overweight. 
In addition, users with a lower weight use more words regarding biological process, (\emph{Tweet \_LIWC\_bio}), such as ``eat'', or ``body'', than their heavier counterparts. Previous research shows that successful weight management is linked to health awareness \footnote{\url{http://www.health.harvard.edu/exercise-and-fitness/lose-weight-and-keep-it-off}}, which matches our findings. 
We observed a number of other top indicators (such as for the categories ppron, social or affect.), but these are admittedly hard to interpret. We hope that our observations help other researchers to form hypotheses around these to test in more depth.

\begin{table}[]
\centering
\begin{tabular}{|l|c|c|c|c|}
\hline
                             &                   & R    & MAE   & RMSE  \\ \hline
                             & Constant Baseline          & -0.14 & 27.89 & 34.67 \\ \hline
\multirow{4}{*}{\begin{tabular}{@{}l@{}}Tweet \\ Only \end{tabular}}  & Language Split    & 0.23 & 23.73 & 30.06 \\ \cline{2-5} 
                             & Gaussian Process  & 0.50 & 23.67 & 29.87 \\ \cline{2-5} 
                             & Gaussian Process + BoW  & 0.55 & 23.00 & 29.04 \\ \cline{2-5}
                             & SVM (Linear Kernel) & 0.34 & 26.65 & 33.31 \\ \hline
\multirow{4}{*}{\begin{tabular}{@{}l@{}}Tweet \\ + \\ Bio \end{tabular}} & Language Split    & 0.07 & 40.99 & 62.32 \\ \cline{2-5} 
                             & Gaussian Process  & 0.48 & 23.84 & 30.26 \\ \cline{2-5} 
                             & Gaussian Process + BoW  & 0.55 & 22.96 & 28.99 \\ \cline{2-5} 
                             & SVM (Linear Kernel) & 0.34 & 26.81 & 33.47 \\ \hline
\end{tabular}
\caption{Results of predicting a person's average weight (in lb) using social media information.}
\label{Performance}
\up
\end{table}

Table~\ref{Performance} shows the weight prediction performance. We evaluate the model performance by three measures: correlation coefficient (R), Mean Absolute Error (MAE) and Root Mean Squared Error (RMSE). The MAE represents an average of the absolute errors; and the RMSE shows the standard deviation of errors. They are evaluated for 391 users using 10-fold cross validation. Specifically, the baseline is shown model performance without Tweet and Bio features; and the language split model is built by splitting the data by languages (English V.S. Japanese).

Initially, using only information from normal tweets, the Gaussian Process model explains around 25\% of the variance in average weight. This indicates that there is a link that's worth exploring between QS data and social media data. Furthermore, by adding their self-description data, the performance (Table~\ref{Performance}) drops a little.

\section{Using Quantified Self Data at the Population Level}
So far, all of our analysis has linked QS and social media data at the individual level. Here, we look at population-level patterns that can be obtained by using the QS weight information obtained through Twitter.  Specifically, we explore the patterns of quantified-self data across days-of-weeks or months-of-years on a larger dataset, including 897 users.

\subsection{Trends in QS Behavior Across Days-of-Week}

\begin{table}[]
\centering
\begin{tabular}{|l|l|l|l|l|l|l|l|l|}
\hline
&           & Mon   & Tue   & Wed   & Thu   & Fri   & Sat   & Sun   \\ \hline
&\begin{tabular}{@{}l@{}}Withings \\ weigh-ins \end{tabular}   & 54k & 54k & 54k & 53k & 47k & 63k & 52k \\ \hline
&\begin{tabular}{@{}l@{}}Fitness\\ tweets \end{tabular}& 9.4k  & 9.6k  & 9.3k  & 9.2k  & 8.9k  & 8.8k & 9.0k  \\ \hline
\multirow{3}{*}{\begin{sideways}Google\end{sideways}}&``bmi'' & 27.1 & 28.5 & 28.2 & 27.8 & 25.0 & 21.6 & 22.8 \\ \cline{2-9}
&\begin{tabular}{@{}l@{}} ``Weight \\ loss'' \end{tabular} & 36.1 & 35.9 & 34.8 & 33.4 & 31.5 & 31.5 & 34.1 \\ \cline{2-9}
&``diet'' & 92.0 & 90.5 & 87.7 & 85.0 & 78.5 & 78.3 & 86.9 \\ \hline
\end{tabular}
\caption{Quantified-Self and Google search behavior on different weekdays.}
\label{SocialBehaviorWeekdays}
\up
\end{table}

Table~\ref{SocialBehaviorWeekdays} reports the frequency of automatic generated weigh-in and fitness tweets in our dataset. For comparison, we also show information for Google search volumes\footnote{\url{https://www.google.com/trends/explore}} for the three queries ``BMI'', ``weight loss'' and ``diet'', summarized for days-of-week from September 2015 to November 2015. 

There are clear weekly patterns detectable for both the QS and the Google Trends data. Put simply, users are most aware of their weight on Saturdays with, by far, the largest number of weigh-ins. However, this is also the day where they are least likely to ``take action'' as defined by (i) generating fitness tweets or (ii) searching for diet-related information. By contrast, ``corrective action'' seems to be most likely on Mondays. Consistent observations were made by Weber and Achananuparp~\cite{weberpalakorn16psb} who observed that the number of users logging their meals with MyFitnessPal is highest (lowest) on Mondays (Saturdays). Of those users that log their meals, the fraction consuming more than their self-set calorie goals is also lowest (highest) on Mondays (Saturdays). Based on this one could say that Saturday is everyone's ``cheat day''.

\subsection{Trends in QS Behavior Across Months-of-Year}

\begin{figure}
  \centering
  \includegraphics[width=1\columnwidth]{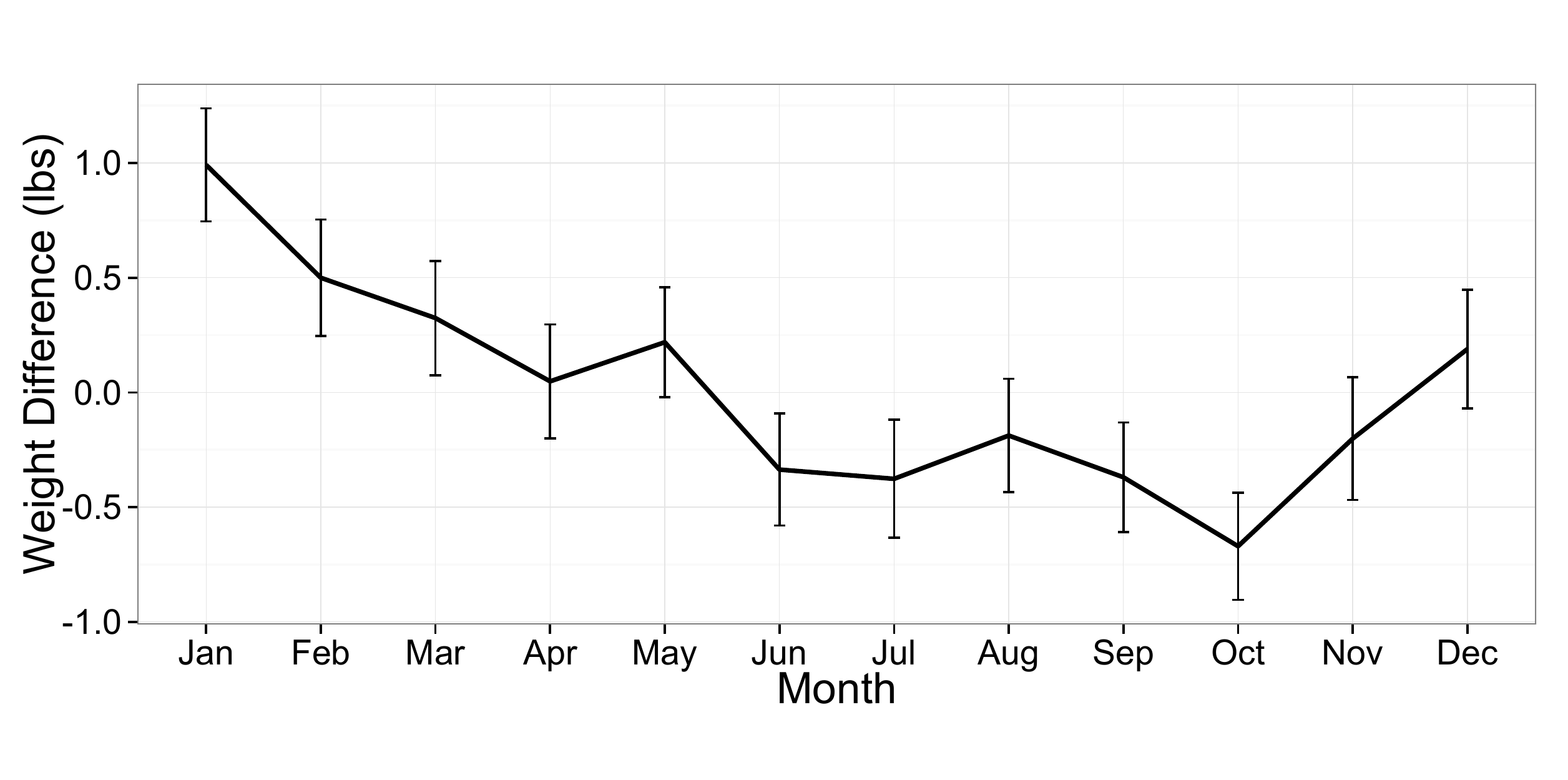}
  \caption{Weight changes for different months,  relative to a user's mean weight.}~\label{fig:WeightChangeMonth}
  \up
  \vspace{-5pt}
\end{figure}

Whereas the previous section looked at weigh-in trends across a week, we look at actual \emph{weight} changes across a year. For each person, we compare their average weight within a given month to their global average weight. For each month, these weight changes are then averaged across all users. Mean and error bar of weight changes for different months are shown in Figure~\ref{fig:WeightChangeMonth}. Monthly patterns are observed showing that users gain weight during winter, from a low in October to a high in January, before starting to lose weight again. Though the observed jump of just under 1lb during the holiday seasons might appear lower than intuition would suggest, this value is perfectly in line with the results of a meta analysis of weight gain over Christmas  \cite{vorland11}.\footnote{Also see \url{http://tinyurl.com/z7r4c5s} for more information on this topic.}

\begin{figure}
  \centering
  \includegraphics[width=1\columnwidth]{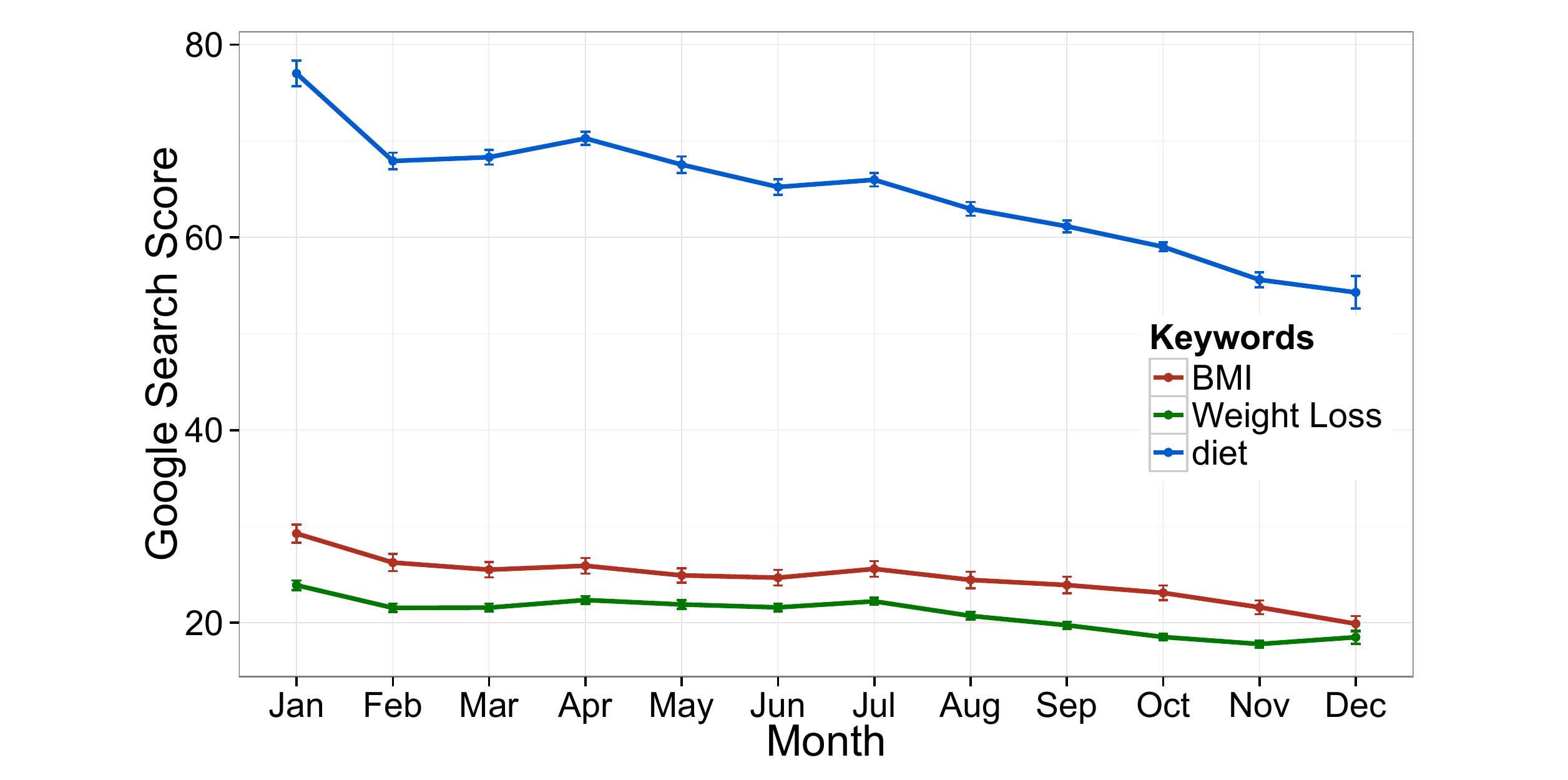}
  \caption{Google Trends search scores for different months from 2005 to 2015.}~\label{fig:GoogleScoreMonth}
  \up
\end{figure}

Similar to our week-based analysis, we wanted to see if there is a link between QS data and topical interest as observed through Google Trends. Figure~\ref{fig:GoogleScoreMonth} presents the mean and the error bar of Google search scores aggregated by month from 2005 to 2015. For all three of our search terms, search activity is highest in January, possibly due to New Year's resolutions. Overall, the search volume changes more abruptly from December to January than the actual weight (see Figure~\ref{fig:WeightChangeMonth}). So whereas users slowly put on pounds from October to January, there appears to be a sudden change in weight loss intent from December to January -- assuming that the selected Google search terms do indeed measure ``weight loss intent''.

\section{Discussion and Limitations}


The data analyzed for this study does not come from a randomized trial or from a representative sample of the population. Users who choose to publicly tweet their weight are likely to differ from a ``normal'' user trying to loose weight, though (i) our population's average weight and (ii) the weight gain over Christmas were surprisingly close to known values. Weber and Mejova \cite{webermejova16dh} show that, with a certain amount of noise, a user's body weight or at least classes such as ``overweight or not'' can be inferred from their Twitter profile pictures. We are not relying on such noisy labels but, basically, we are trading a loss in recall for an increase in precision.

A considerable fraction of users, 198 out of 391, had chosen Japanese as their interface language and, correspondingly, many of their tweets were not be in English. Google's automatic translation might introduce errors, though for tasks such as sentiment analysis machine translation typically performs sufficiently well \cite{baneaetal08emnlp}.

\section{Conclusions}

In this paper, we presented a study that combines quantified self data from internet-enabled smart scales with general social media data on Twitter. We used this combination of data sources to predict a user's weight using only their social media activity.  
Our data also capture weekly patterns, such as a peak of weigh-in activity on Saturday, and monthly patterns, such as a weight increase over Christmas. 
We believe that such a data fusion between messy, general life style social media data and very accurate, longitudinal quantified self data has great potential to improve personalized health care.

\thanks{This is a preprint of an article appearing at ACM DigitalHealth 2016.}

\bibliographystyle{abbrv}
\bibliography{sigproc}  

\end{document}